\documentclass{article}

\usepackage{arxiv}

\usepackage[utf8]{inputenc} 
\usepackage[T1]{fontenc}    
\usepackage{hyperref}       
\usepackage{url}            
\usepackage{booktabs}       
\usepackage{amsfonts}       
\usepackage{nicefrac}       
\usepackage{microtype}      
\usepackage{lipsum}		
\usepackage{graphicx}
\usepackage{natbib}
\usepackage{doi}
\usepackage{enumitem}
\usepackage{cite}
\usepackage{amsmath,amssymb,amsfonts}
\usepackage{algorithm}
\usepackage{algorithmic}
\usepackage[dvipsnames]{xcolor}
\usepackage{graphicx}
\usepackage{caption}
\usepackage{textcomp}
\usepackage{tabularx}
\usepackage{graphicx}
\usepackage{url}
\usepackage{amsmath}
\usepackage[dvipsnames]{xcolor}
\usepackage{soul}
\definecolor{purple(html/css)}{rgb}{0.5, 0.0, 0.5}
\usepackage{enumitem}

\definecolor{purple(html/css)}{rgb}{0.5, 0.0, 0.5}
\usepackage{enumitem}
\definecolor{goldenpoppy}{rgb}{0.99, 0.76, 0.0}
\definecolor{darkslategray}{rgb}{0.18, 0.31, 0.31}
\definecolor{black}{rgb}{0.0, 0.0, 0.0}
\definecolor{darkpastelgreen}{rgb}{0.01, 0.75, 0.24}
\definecolor{orange(webcolor)}{rgb}{1.0, 0.65, 0.0}
\definecolor{darkorange}{rgb}{1.0, 0.55, 0.0}
\definecolor{orange(ryb)}{rgb}{0.98, 0.6, 0.01}
\definecolor{orange-red}{rgb}{1.0, 0.27, 0.0}
\definecolor{deepcarrotorange}{rgb}{0.91, 0.41, 0.17}
\definecolor{lightblue}{rgb}{0.68, 0.85, 0.9}
\definecolor{blue(ncs)}{rgb}{0.0, 0.53, 0.74}

\title{Organ localisation using supervised and semi supervised approaches combining reinforcement learning with imitation learning}

\author{ \href{https://orcid.org/0000-0001-7580-5602}{\includegraphics[scale=0.06]{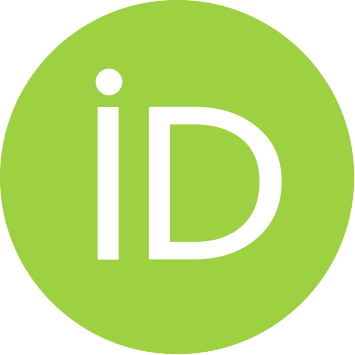}\hspace{1mm}Sankaran Iyer}\\
	School of Computer Science and Engineering,
	University of New South Wales
	Kensington, NSW 2033 \\
	\texttt{sankaran.iyer@unsw.edu.au} 
	\AND
	Alan Blair \\
	School of Computer Science and Engineering,
	University of New South Wales
	Kensington, NSW 2033 \\
	\texttt{blair@cse.unsw.edu.au} 
	\And
	Laughlin Dawes \\
	Department of Medical Imaging, Prince of Wales Hospital, NSW, Australia \\
	\texttt{Laughlin.Dawes@health.nsw.gov.au} 
	\And
	Daniel Moses \\
	Department of Medical Imaging, Prince of Wales Hospital, NSW, Australia \\
	\texttt{daniel.moses@health.nsw.gov.au} 
	\And
	Christopher White \\
	Department of Endocrinology and Metabolism, Prince of Wales Hospital, NSW, Australia \\
	\texttt{christopher.white@health.nsw.gov.au} 
	\And
	Arcot Sowmya \\
	School of Computer Science and Engineering,
	University of New South Wales
	Kensington, NSW 2033 \\
	\texttt{a.sowmya@unsw.edu.au} 
}

\renewcommand{\headeright}{}
\renewcommand{\shorttitle}{}

\hypersetup{
pdftitle={A template for the arxiv style},
pdfsubject={q-bio.NC, q-bio.QM},
pdfauthor={David S.~Hippocampus, Elias D.~Striatum},
pdfkeywords={First keyword, Second keyword, More},
}

\begin{document}
\maketitle

\begin{abstract}
Computer aided diagnostics often requires analysis of a region of interest (ROI) within  a radiology scan, and the ROI may be an organ or a suborgan. Although deep learning algorithms  have the ability to outperform other methods, they  rely on the availability of a large amount of annotated data. Motivated by the need to address this limitation, an approach to localisation and detection of multiple organs  based on supervised and semi-supervised learning is presented here. It draws upon previous work by the authors on localising the thoracic and lumbar spine region in CT images. The method generates six bounding boxes of organs of interest, which are then fused to a single bounding box. The results of  experiments on localisation of the Spleen, Left and Right Kidneys in CT Images using  supervised and semi supervised learning (SSL) demonstrate the ability  to address data limitations with a much smaller data set and fewer annotations, compared to other state-of-the-art methods. The SSL performance was evaluated using three different mixes of labelled and unlabelled data (i.e.30:70,35:65,40:60) for each of lumbar spine, spleen left and right kidneys respectively. The results indicate that  SSL provides a workable alternative especially in medical imaging where it is difficult to obtain annotated data. 
\end{abstract}


\section{Introduction}
\label{sec:introduction}
Organ/region of interest (ROI) localisation in a radiology scan is an important task in medical image analysis. It helps to focus attention on a specific area or volume for further analysis and processing. It can lead to better segmentation by reducing false positives and improved accuracy in lesion detection by reducing the search space. A fully automated computer aided localisation method is beneficial because manual localisation is resource intensive.  Also,  3D localisation is important in light of the increasing use of 3D data sets including Computed Tomography (CT) and Magnetic Resonance Imaging (MRI). 
Processing of 3D data sets  continues to be a challenge due to the need to handle variations in  organ shape, size, field of vision and pathology. The localisation method described here extends a previously reported localisation method for thoracic and lumbar spine localisation \citep{Iyer2020}, \citep{Iyer2019}to handle limited annotated CT datasest in both supervised and  semi-supervised setting. The algorithm generates six bounding boxes around the organ / ROI by combining deep reinforcement learning with imitation learning. A novel method to combine the  multiple bounding boxes generated by the algorithm has also been developed. Besides Jaccard index and Dice Coefficient, average wall to wall and centroid to centroid distances in comparison with the ground truth are also reported. 

The target organs are the spleen, left and right kidneys besides lumbar spine  in  abdominal CT scans. The organs were chosen to demonstrate the ability of the algorithm to localise any organ/ROI. It is important to note however that surgeons may be interested in any region, and not necessarily  just  anatomical ones. Deep learning algorithms with their proven ability outperform other methods, however they rely on a significant amount of data being available,  annotation can be expensive and it is not always possible to apply fully supervised methods for these reasons. It is often a challenge to deal with limited annotations and there has been  growing demand to address this issue using other techniques such as semi supervised learning (SSL). Hence in this work, the localisation method is extended work in a semi-supervised setting.

The proposed localisation method has shown the ability to address these issues on two fronts:
\begin{enumerate}[label=\alph*)]
    \item perform in supervised mode with a limited data set
    \item perform in SSL mode by  striking a balance between limited labelled data and a larger amount of unannotated data. 
\end{enumerate}
Experiments were conducted with three different mixes of labelled and unlabelled data. The results demonstrate the promise and feasibility of the approach. Based on published literature, this is the first attempt to employ SSL for organ localisation.
The rest of the paper is organised as follows: related work is reviewed in section \ref{sec:relatedWork},  background and proposed work are described in section 3, the proposed methods are presented in section 4, experiments and results in section 5 and conclusion in section 6.
\section{Related Work}
\label{sec:relatedWork}
\subsection{Localisation algorithms}
\label{localg}
A variety of methods have been proposed for automatic organ localisation, which rely on multi-atlas registration or machine learning using hand-crafted features \citep{Criminisi2013}, \citep{Jimenez2016}, \citep{Zhou2012}. \citep{Zheng2009}. These methods can be computationally intensive or highly dependent on feature selection. 
Recent methods make use of deep learning with their ability to automatically extract features. Some methods perform landmark detection by combining the detected organs in 2D slices in the axial, coronal and sagittal planes to estimate 3D bounding boxes \citep{Hussain2017},\citep{deVos2017},\citep{Lu2016}.  Region-proposal based localisation \citep{Shaoqing2017} has shown a lot of promise in object detection and has been extended to multi organ detection \citep{Xu2019}. One of the salient aspects of region proposal based methods is the generation of multiple candidate bounding boxes that may overlap with each other. They use "non maximum suppression" to reduce the redundant bounding boxes. Xu \textit{et al} used a different technique, whereby they fused the bounding boxes using the weighted mean of candidate bounding boxes based on class scores. The approach used in this paper is motivated by this idea of generating candidate bounding boxes, predicting their confidence scores and fusing them into one.

Deep reinforcement learning has been used for landmark detection \citep{Ghesu2017} \citep{Alansary2019}, and  extended to bounding box localisation \citep{Navarro2020}. The localisation strategy used in this work is inspired by localisation using deep reinforcement learning, with random search replaced by guided search using imitation learning.

\subsection{Deep Semi supervised learning (SSL)}
The state of the art in deep SSL methods has been summarised by \citep{Ouali2020} and \citep{Cheplygina2019}. These algorithms are applied in situations wherein there is a  limited number of annotated samples together with a much larger number of unlabelled samples. The objective is to train a model that leverages the available labelled samples. There are broadly four categories of SSL methods: consistency regularisation, proxy labelling, generative models and graph based methods. There is very little work on organ localisation that employ SSL.   "Focal Mix" has been used \citep{Wang2020} which is based on "MixMatch", a holistic approach combining all the categories mentioned above. An altogether different approach was taken by \citep{LiZ2019} for thoracic disease localisation from X Rays composed in part of bounding box annotations for diseases and just disease annotations for the rest. They trained a ResNet model with a loss function that combines bounding loss if annotations are available, and simply the probability based on image level disease annotations for the rest. 

One popular SSL approach in medical imaging is self training, which falls under the proxy labelling category. This involves building a model with limited training samples and using it to predict labels of the unlabelled samples. The predicted labels or a subset of the predicted labels are then added to the annotations as "pseudo labels" and a new model is retrained with the new annotations. The process of predicting labels and retraining models with the new labels is repeated several times. There are variations to this approach in how the predicted labels are added. Some works introduce a method of determining a confidence level that exceeds a user defined threshold \citep{Cheplygina2019}.

The primary motivation to use the semi supervised learning approach is driven by the fact that obtaining annotated samples is often a challenge in medical imaging studies. Therefore self training or pseudo labelling has been experimented with to explore the feasibility of bounding box localisation of anatomical parts.

\subsection{Contributions}
The following are the major contributions of this work:
\begin{enumerate}[label=\alph*)]
    \item extension of previously reported method for thoracic and lumbar spine localisation that combines reinforcement learning with imitation learning, to localisation of left and right kidneys and spleen, generating six candidate bounding boxes.
    \item a novel method to fuse multiple bounding boxes, with performance reported using other popular metrics suited to the task, including mean wall to wall and centroid to centroid distances
    \item adapting self training, a semi supervised learning approach, to localisation of the lumbar spine, left and right kidneys and spleen, and comparing the performances with the fully supervised approach.
\end{enumerate}
\section{Proposed Method}
\label{method}
\subsection{Ethics Approval}
\label{sec:ethics}
This work was performed  after getting approval no. HC200485 from the Human Research Ethics Committee (HREC) of  UNSW. This HREC is constituted and operates in accordance with the  Australian National Health and Medical
Research Council’s (NHMRC)  National Statement on Ethical Conduct in Human Research
(2007). The processes used by this HREC to review multi-centre research proposals have been
certified by the National Health and Medical Research Council.
\subsection{Overview}
\label{sec:summary}
The proposed method to achieve 3D organ localisation in CT images combines multiple bounding boxes generated by deep reinforcement learning and imitation learning. This is an extension of previous work  on thoracic and lumbar spine localisation \citep{Iyer2020} and \citep{Iyer2019} to the localisation of spleen, lumbar spine, left and right kidneys in a semi-supervised setting. The method involves generating 6 bounding boxes using three different CNN architectures as detailed in \ref{backgound}. The proposed method may be used to localise other regions of interest too, and  not necessarily organs.

As stated in section \ref{localg}, the idea of generating multiple target bounding boxes and fusing them is similar to the modification of Faster RCNN  for 3D organ localisation by \citep{Xu2019}. The method used in this work involves generating six bounding boxes and their fusion is achieved by using a novel method to generate confidence scores associated with candidate bounding boxes as described in section \ref{BBFus}.
The SSL technique used in this work is self training, in which pseudo labels are added to the training set, provided the predicted confidence value exceeds a threshold level. The localisation algorithm is suitably adapted to train a model using this approach and results are reported on 3 different mixes of labelled/unlabelled data. As a baseline for comparison, a fully supervised model is also trained and tested.
\subsubsection{Background}
\label{backgound}
Earlier works by the authors \citep{Iyer2019} and \citep{Iyer2020} cover the necessary background behind the proposed approach. The localisation method is a guided approach based on imitation learning that uses a navigation strategy to locate a landmark as illustrated by Figure 1 of previous work \citep{Iyer2020}. The strategy involves navigating in the coordinate direction that is at maximum distance from the current location to the centre of the ground truth. This is in contrast to the random search approach of reinforcement learning. The approach adopted for 3D bounding box localisation combines the deep Q learning algorithm \citep{LiY2017} with imitation learning for locating a target ROI from a predefined starting point in the image. It utilises two different networks during the search process. The first is a policy network trained using imitation learning to guide a pre-selected bounding box to the target. The second is a bounding box network, which is a regression model to predict the dimensions of bounding boxes. The search method used in the earlier works was primarily targeted at localising thoracic and lumbar spine regions  in CT images of 3mm and 5mm slice thickness, as a preliminary step in extracting ROIs for vertebral compression fracture detection.
The work in this paper is more concerned with establishing the generalisability of the algorithm to localise other organs/ROIs. It is also intended to test the ability to achieve good results with limited data as well as perform in SSL mode.
\subsubsection{Annotation}
\label{annotation}
\begin{figure}[t!]
    \includegraphics{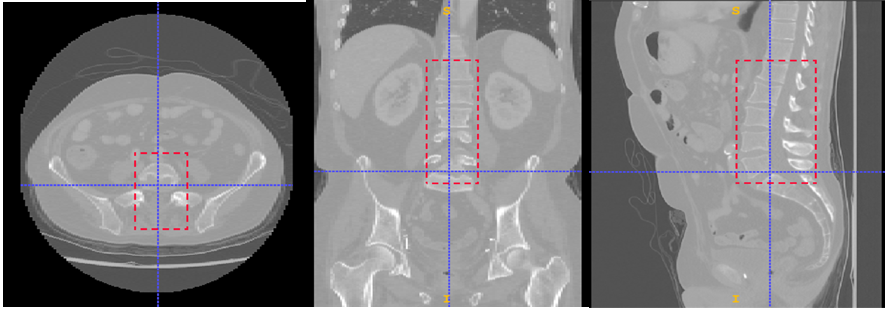}
    \centering
    \caption{\centering{3D Bounding Box annotation using ITK-SNAP shown by white boxes in 3 planes surrounding a lumbar spine}}
    \label{fig:Annotation}
\end{figure}

\begin{figure}[t!]
    \includegraphics[scale = 1.2]{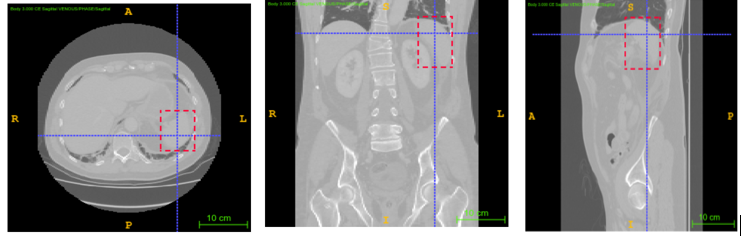}
    \centering
    \caption{\centering{Spleen Annotation}}
    \label{fig:spleen}
\end{figure}

\begin{figure}[t!]
    \includegraphics[scale = 1.3]{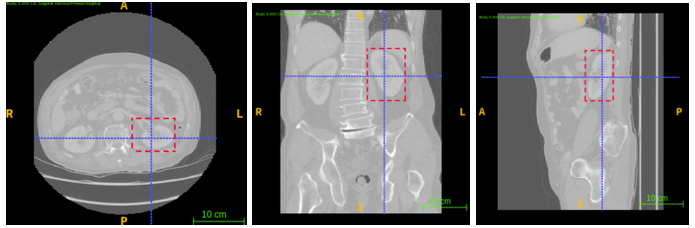}
    \centering
    \caption{\centering{Left Kidney Annotation}}
    \label{fig:left Kidney}
\end{figure}

\begin{figure}[t!]
    \includegraphics[scale = 1]{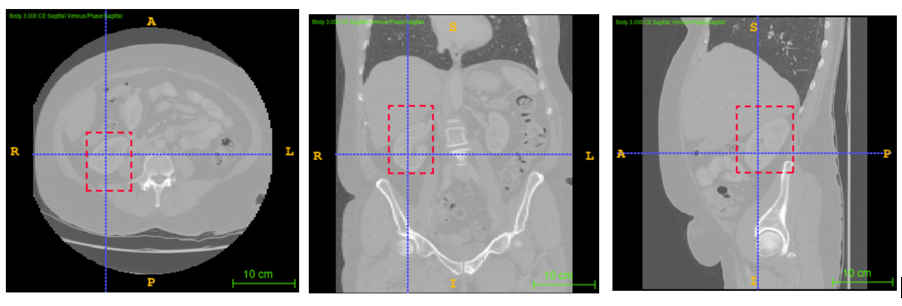}
    \centering
    \caption{\centering{Right Kidney Annotation}}
    \label{fig:right Kidney}
\end{figure}

\hfill\\
The annotation process using ITK-SNAP in the three planes is illustrated in Fig \ref{fig:Annotation} for creating a 3D bounding box around a lumbar spine region.
The annotations of spleen, left and right kidneys are illustrated by Figures \ref{fig:spleen}, \ref{fig:left Kidney} and \ref{fig:right Kidney} respectively.

\subsubsection{Performance Measures}
\label{sec:perfmeasures}
\hfill\\
The performance of the algorithm is evaluated using the following metrics:
\begin{enumerate}[label=\alph*)]
\item \textbf{Intersection over Union (IOU)} of the predicted bounding box with the ground truth box. IOU is also known as \textbf{Jaccard Index}.
\item \textbf{Dice Coefficient (DC)}, which is the ratio of twice the intersection over sum of the volumes of ground truth and predicted bounding boxes. 
\item The \textbf{centroid distance} between the centroids of the ground truth and predicted bounding boxes. 
\item The average \textbf{wall distance} between the 6 walls of the predicted bounding boxes and the ground truth bounding box.

\end{enumerate}
\subsubsection{Algorithm}
\label{sec:algorithm}
\hfill\\
The algorithm proposed in this paper is a modification of an earlier algorithm by the authors \citep{Iyer2019} and the process is illustrated in Figure 3 of earlier work \citep{Iyer2020}. It involves training of two networks: 
\begin{enumerate}[label=\alph*)]
    \item the policy network navigates a preselected bounding box to the centre of the ROI. 
    \item the bounding box network predicts the actual size of the bounding box surrounding the ROI along with the predicted IOU, which is used as a confidence level attached to each bounding box.
\end{enumerate}
\begin{algorithm}

\caption{Localisation by combining deep reinforcement learning with imitation learning for ROI detection}
\label{alg1}
\textbf{Input:} CT abdominal 3D datasets.

\textbf{Output:}\textit{Policy function} for navigating bounding Box.\newline 
\textit{Bounding Box function} that predicts the actual bounding box coordinate sizes along with confidence level each bounding box\newline
initialize Policy replay memory D\newline
initialize Bounding Box replay memory B\newline
initialize action-value function Q with random weights\newline
\textbf{for} cycles from 1 to M\newline
\hspace*{0.25cm}\textbf{for} each range of starting points \newline
\hspace*{0.5cm}\textbf{for} {each randomly selected scan}\newline
\hspace*{0.75cm}set a pre-selected bounding box at a predefined starting point =  s$_1$\newline
\hspace*{1cm}\textbf{for} steps from 1  to  N \newline
\hspace*{1.25cm}following \textit{$\epsilon$-greedy} policy select an action\newline
\[
a_t = \left\{
            \begin{aligned}
            \textit{Imitation action with probability $\epsilon$}\\
            \textit{argmax$_a$Q(s$_t$,a) otherwise}\\
            \textit{correction is applied if the predicted action is away from target}\\
            \end{aligned}
         \right.
\]\\
\hspace*{1.25cm}store transition s$_t$,a$_t$ in D\newline
\hspace*{1.25cm}calculate the IOU of s$_t$ with the ground truth\newline
\hspace*{1.25cm}\textbf{if} IOU $>=$ threshold\newline
\hspace*{1.5cm} store s$_t$,\textit{ground truth bounding box coordinate} sizes along with calculated IOU in B\newline
\hspace*{1.25cm}set s$_t$=s$_{t+1}$\newline
\hspace*{1.25cm}\textbf{if} bounding box centre = ground truth centre\newline
\hspace*{1.5cm}set a$_t$= “Terminate”\newline
\hspace*{1.5cm}store resulting transitions in D and B\newline
\hspace*{1.25cm}\textbf{break}\newline
\hspace*{1cm}\textbf{end for}\newline
\hspace*{1cm}train Policy network with random samples from D using mean square error loss\newline
\hspace*{1cm}train Bounding Box network with random samples from B using mean square error loss\newline
\hspace*{.75cm}\textbf{end for}\newline
\hspace*{0.5cm}\textbf{end for}\newline
\hspace*{0.25cm}\textbf{end for}\newline
\end{algorithm}
\begin{figure}[t!]
    \includegraphics[scale = 0.8]{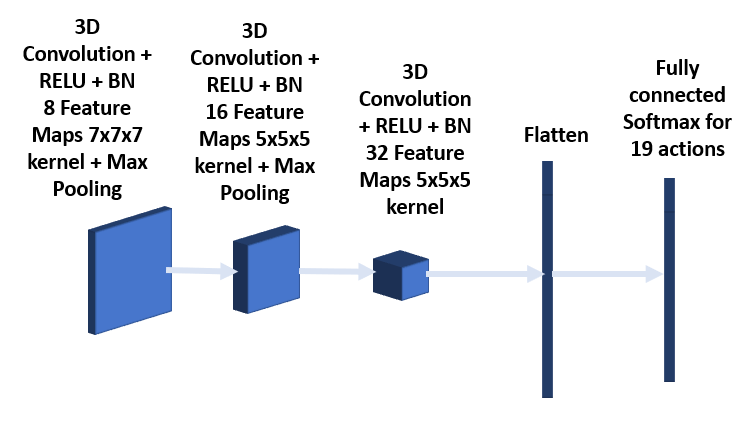}
    \centering
    \caption{CNN architecture 1 for Navigation Network}
    \label{fig:LocNav1}
\end{figure}
\begin{figure}[t!]
    \includegraphics[scale = 0.8]{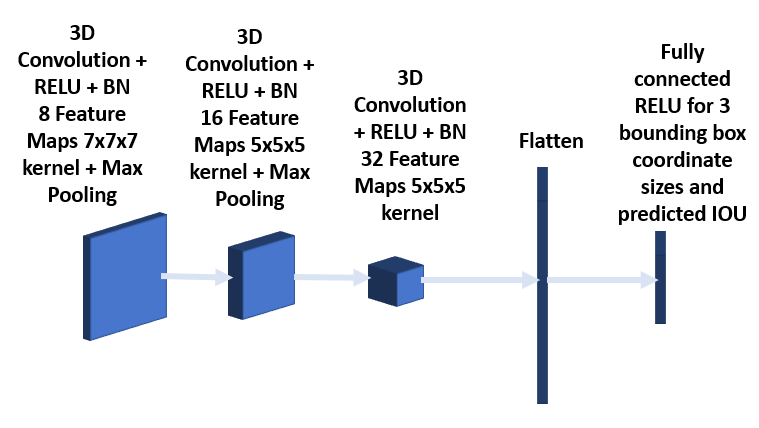}
    \centering
    \caption{CNN architecture 1 for Bounding Box Network}
    \label{fig:LocBB1}
\end{figure}

\begin{figure}[t!]
    \includegraphics[scale = 0.8]{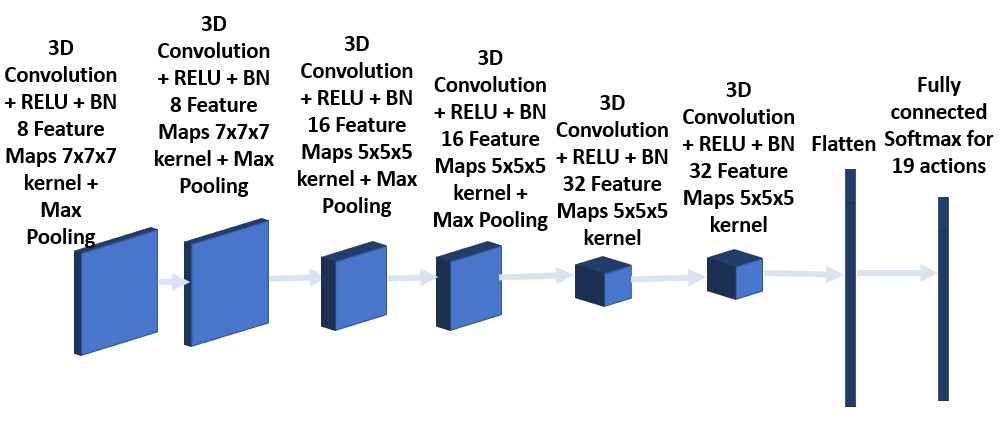}
    \centering
    \caption{CNN architecture 2 for Navigation Network}
    \label{fig:LocNav2}
\end{figure}

\begin{figure}[t!]
    \includegraphics[scale = 0.8]{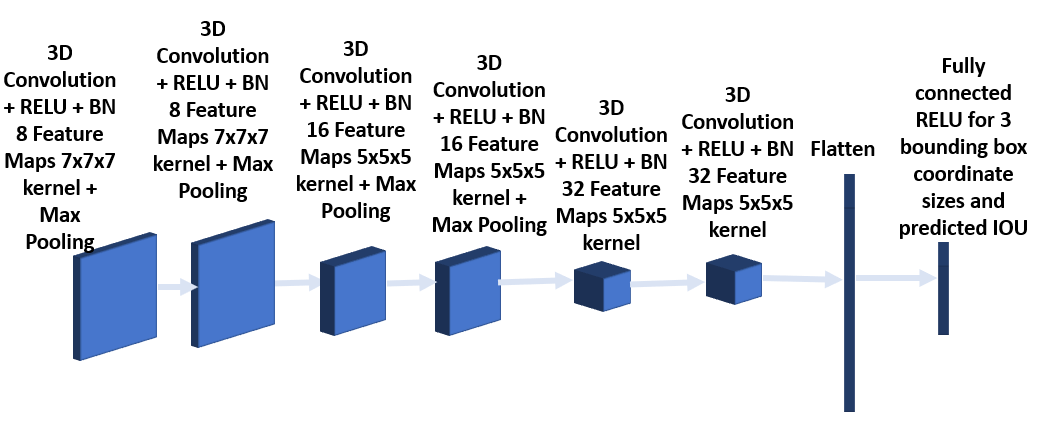}
    \centering
    \caption{CNN architecture 2 for Bounding Box Network}
    \label{fig:LocBB2}
\end{figure}

\begin{figure}[t!]
    \includegraphics[scale = 0.8]{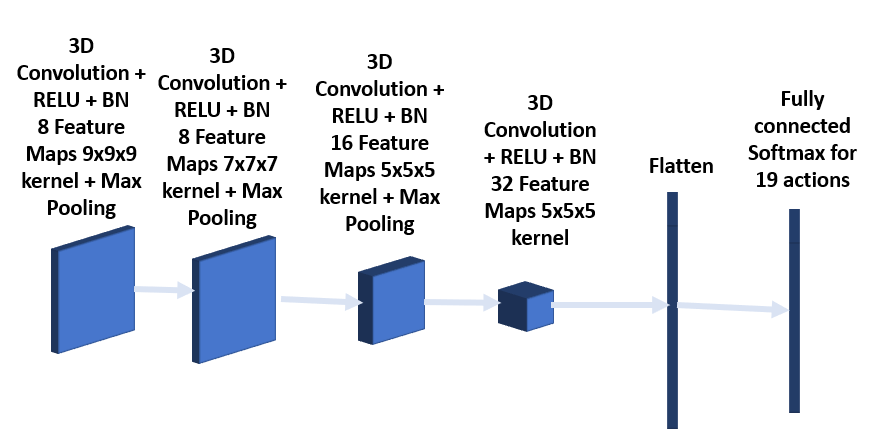}
    \centering
    \caption{CNN architecture 3 for Navigation Network}
    \label{fig:LocNav3}
\end{figure}

\begin{figure}[t!]
    \includegraphics[scale = 0.8]{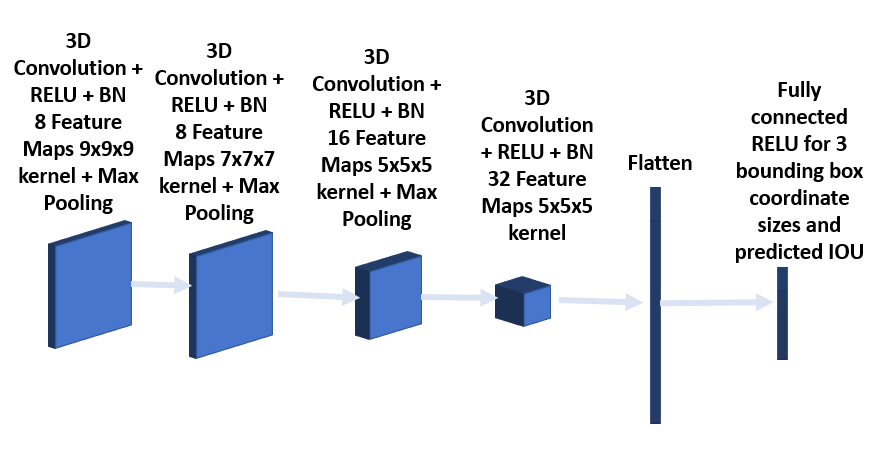}
    \centering
    \caption{CNN architecture 3 for Bounding Box Network}
    \label{fig:LocBB3}
\end{figure}
The modified pseudo code is presented in Algorithm \ref{alg1}. The policy network that is trained to predict the coordinate direction of shift (action) for an image region bounded by an initial pre-selected bounding box. It is  also trained to predict the IOU which is used as an indicator of the confidence level associated with the bounding box. In each coordinate direction, 3 levels of movement of the bounding box in both positive and negative directions require 6 actions. For the three coordinate directions, therefore, 18 actions are possible.

The imitation function guides the navigation process by returning an action, which is the coordinate direction at maximum distance from the ground truth centre. It also performs the necessary corrections to predictions that deviate from the intended course. Depending on the distance between the current and ground truth centres, an appropriate level of movement selected: coarse, fine or very fine. The rest of the training process is the same as discussed in earlier work \citep{Iyer2019} for starting navigation with a pre-selected bounding box from a preset location as well as other starting points to recover from unfamiliar locations.

The policy network shown in Figure \ref{fig:LocNav1}  is made up of three 3D convolution layers together with batch normalization and RELU activation. The kernel sizes of the first, second and third convolution layers are 7x7x7, 5x5x5 and 3x3x3 respectively. The network takes as input the data within the bounding box shrunk by half. The convolution layers are followed first by a fully connected layer and then by a softmax layer for 19 possible actions. The loss function used is mean square error.

The bounding box network shown in Figure \ref{fig:LocBB1} is trained to predict the three coordinate sizes of the ROI. As the pre-selected bounding box is navigated, the regions whose IOUs exceed a threshold level are stored, along with ground truth sizes for training the bounding box network. The predicted IOU from the bounding box network is used as a confidence level associated with the bounding box. This feature is used in bounding box fusion (\ref{BBFus}) and also in semi supervised learning for pseudo label selection.  The bounding box network consists of three 3D convolution layers together with batch normalization and RELU activation. The kernel size of the first, second and third convolution layers are 7x7x7, 5x5x5 and 3x3x3 respectively. The convolution layers are followed first by a fully connected layer and then by a RELU layer for 3 coordinate sizes. The loss function used is mean square error.

To improve overall performance, two other CNN architectures were trained besides the one described above, and the predicted bounding boxes using all three models are provided to the next stage for analysis. The architecture of the second model consists of 6 convolution layers as shown in Figures \ref{fig:LocNav2} and \ref{fig:LocBB2} respectively. The first 2 layers have kernel size 7x7x7, followed by 2 convolution layers with kernel size 5x5x5 and the final 2 convolution layers having kernel size 3x3x3. Each convolution layer is followed by batch normalization. Max pooling is added after the second and fourth layer. The third model as shown in Figures \ref{fig:LocNav3} and \ref{fig:LocBB3} has a convolution layer with 9x9x9 kernel and a batch normalization preceding the architecture in the model shown in Figures \ref{fig:LocNav1} and \ref{fig:LocBB1}.
\subsubsection{Testing}
\hfill\\
\label{testing}
In the testing mode there is no imitation learning involved. The search is terminated after 25 steps, or when a  “Terminate” action is triggered, or when a loop is detected between the states. The threshold level for successful detection was set at DC 50\%. The bounding box prediction was run on all the steps and two different methods were used to predict its size: (a) the predicted size in the terminating state and (b) the mean size of the predicted bounding boxes in the last 10 states. This gives rise to 2 bounding box predictions for each CNN model, therefore 6 for the three CNN models used. After analysing performances it was decided to retain both the methods as their performances were similar and there were advantages in retaining individual performances.

\begin{figure}[t!]
    \includegraphics{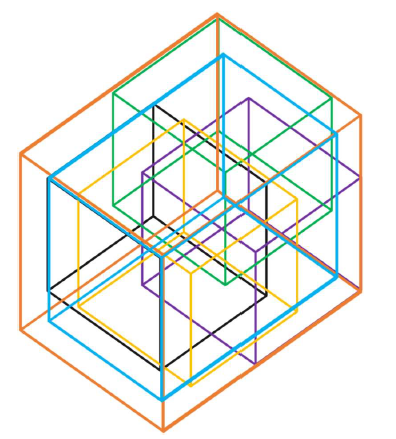}
    \centering
    \caption{Bounding box Fusion generated from 4 candidate bounding boxes shown in \textbf{\textcolor{black}{black}}, \textbf{\textcolor{goldenpoppy}{gold}}, \textbf{\textcolor{darkpastelgreen}{green}} and \textbf{\textcolor{purple(html/css)}{purple}} colours. The bounding boxes are fused together to generate the outermost \textcolor{deepcarrotorange}{orange} coloured box which encompasses all individual bounding boxes. This bounding box is fine-tuned, shown in \textbf{\textcolor{blue(ncs)}{blue}}, by offsetting the bounds towards the centroid by a factor arrived at by experiments}
    \label{fig:BBfusion}
\end{figure}
\subsubsection{Bounding Box Fusion}
\label{BBFus}
\hfill\\
As explained in section \ref{testing}, 6 bounding boxes are identified from 3 CNN architectures. There are two options available to process the bounding boxes. Each of the six different bounding boxes may be processed independently and the results combined using ensembling methods. The other option is to combine them into a single bounding box.
As discussed in section \ref{sec:algorithm} and Algorithm \ref{alg1}, the Bounding Box network is trained to predict the confidence value associated with its prediction by training with the IOU calculated for the bounding box as it approaches the ground truth centroid. 

Algorithm \ref{alg2} is used to fuse the bounding boxes,  resulting in a new bounding box that encompasses all the bounding boxes. The IOU of the resulting bounding box is likely to be lower than the best performing bounding box, however it will contain all the relevant information captured by the individual bounding boxes. The bounds are further fine-tuned by adding an offset factor 'O' that is a percentage of the difference between the lowest and highest values of the lower bound of the bounding boxes. This factor is decided based on the experimental results of five fold cross validation, which are discussed in section \ref{fusoffst}.
\subsection{Semi Supervised Learning by Self Training}
\label{selftraining}
As introduced in section \ref{sec:summary}, the semi supervised learning approach used in this work is Self Training with pseudo labels added if the confidence level exceeds a threshold level. The pseudo code for Self Training is presented in Algorithm \ref{alg3}. It makes use of the confidence level of bounding boxes predicted using Algorithm \ref{alg1}. The fused bounding box coordinates of Algorithm \ref{alg2} are used as pseudo bounding box labels if the criterion of confidence values is achieved.
\setlength{\textfloatsep}{0pt}
\begin{algorithm}
\caption{Bounding Box Fusion}
\label{alg2}
\textbf{Input:} 6 bounding boxes and associated confidence scores\newline
\textbf{Output:} Single fused bounding box\newline 
initialise the list of target bounding boxes to be fused together in $T$,\\
lower bounds of coordinates $L$,\\
upper bound of coordinates $U$\\
Offset factor for fusion $O$\\
\textbf{for} each CT scan in Test set\\
\hspace*{0.25cm}\textbf{for} the confidence values associated with the 2 bounding boxes from each CNN architecture\\
\hspace*{0.50cm} take the sum the confidence values\\
\hspace*{0.50cm} select the CNN architecture which has generated the highest sum and add the bounding boxes to $T$\\
\hspace*{0.50cm} \textbf{if} the sum of the confidence values of each of the remaining CNN architectures exceed threshold \\
\hspace*{0.75cm} Add the 2 bounding boxes generated by the respective CNN architectures to $T$ \\
\hspace*{0.50cm} \textbf{for} each bounding box in $T$\\
\hspace*{0.75cm} identify the lower bound of the three coordinates and store in $L$ \\
\hspace*{0.75cm} identify the upper bound of the three coordinates and store in $U$ \\
\hspace*{0.5cm} take the lowest value of the lower bound $L$\\ \hspace*{0.75cm}add to it an offset factor which is $O$\% between the lowest and highest value in $L$\\
\hspace*{0.5cm}this will for the lowest bound of the fused bounding box\\
\hspace*{0.5cm} take the highest value of the upper bound $U$\\
\hspace*{0.75cm}subtract from it  an offset factor which is $O$\% between the lowest and highest value in $U$\\
\hspace*{0.5cm} this will for the highest bound of the fused bounding box\\
\end{algorithm}
\begin{algorithm}
\caption{Semi Supervised Learning: Self Training}
\label{alg3}
\textbf{Input:} CT abdominal 3D data sets part of which is labelled\\
\hspace*{0.25cm} with bounding box coordinates and the remaining unlabelled\\
\textbf{Output:}\textit{Policy function} for navigating bounding Box.\newline 
\hspace*{0.25cm}\textit{Bounding Box function} that predicts the actual bounding box \\
\hspace*{0.5cm}coordinate sizes along with confidence level each \\
\hspace*{0.5cm}bounding box\newline 
\textbf{Pre-train} model with labelled set with annotated bounding\\
\hspace*{0.25cm}box coordinates \\
\textbf{Repeat until optimal performance on validation set is reached}\\
\hspace*{0.25cm}\textbf{Test} using Algorithm \ref{alg1} on unlabelled samples and predict \\
\hspace*{0.5cm}coordinates and confidence level for each of the 6 bounding\\
\hspace*{0.5cm}boxes from 3 CNN models\\
\hspace*{0.25cm}\textbf{for} the results of each unlabelled sample\\
\hspace*{0.5cm}\textbf{if} the sum of the confidence values of 2 bounding boxes\\
\hspace*{0.75cm}of any of the three CNN architectures exceed threshold\\
\hspace*{0.75cm}value \\
\hspace*{0.75cm}\textbf{run} bounding box fusion using Algorithm \ref{alg2}\\ 
\hspace*{1cm} and add the unlabelled sample to the labelled with\\
\hspace*{1cm}fused coordinates as pseudo bounding box labels\\
\hspace*{0.25cm} \textbf{retrain} the model with new labelled samples\\
\hspace*{0.5cm}that includes pseudo bounding box labels
\end{algorithm}
\section{Experiments and Results}
Training was performed using Keras/Tensorflow and each cycle of training was run for 20 epochs for supervised learning. Semi supervised learning required running the training for multiple cycles.
\subsection{Dataset}
The dataset was provided by the Prince of Wales Hospital, Randwick, NSW, Australia in an anonymised form after ethics approval (refer \ref{sec:ethics}), and included 110 abdomen CT scans with a slice thickness of 3mm for  lumbar spine and left kidney and 104 scans for spleen and right kidney localisation. 
The data set was manually annotated and verified by a radiologist with 10 years of experience, to identify the two diagonally opposite corner points of a 3D bounding box around the ROI. Annotations were performed to localise left and right kidneys using ITK-SNAP as described in section \ref{annotation}.
\subsection{Supervised Learning with Bounding Box Fusion}
\label{SupResults}
Two sets of experiments were conducted in supervised training mode:

\begin{enumerate}[label=(\alph*)]
    \item five fold cross validation
    \item three fold cross validation to establish a baseline for comparison with experiments performed in semi supervised learning mode
\end{enumerate}
\subsubsection{Offset for bounding box fusion}
\label{fusoffst}
\begin{figure}[t!]
    \includegraphics{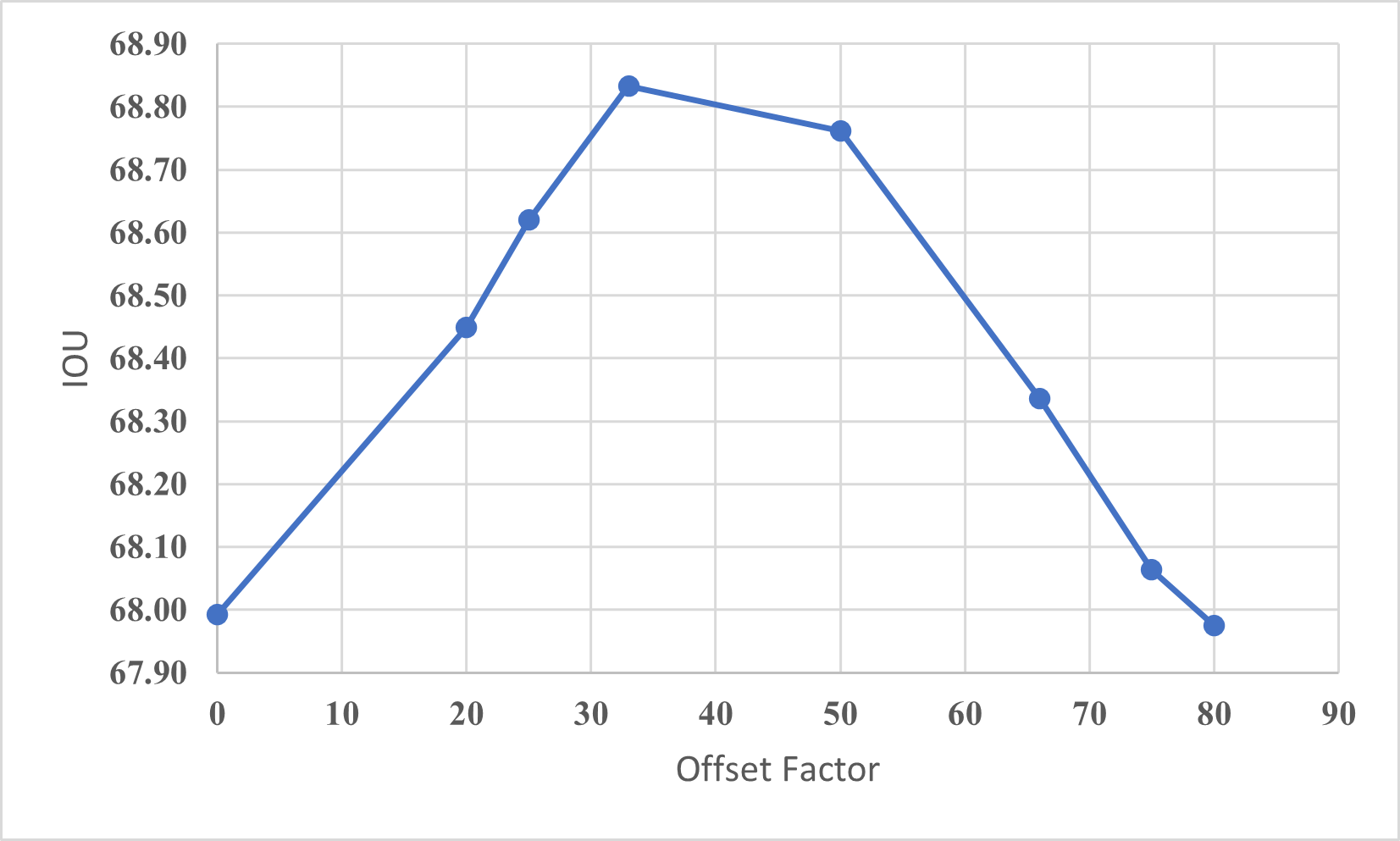}
    \centering
    \caption{Plot of average IOU against Offset factor for Left and Right Kidneys.  Optimal performance is reached when the offset value is 33\%}
    \label{fig:iouplot}
\end{figure}

\hfill\\
Following the procedure outlined in section \ref{BBFus}, the candidate bounding boxes were fused together to generate an all-encompassing bounding box  shown in \textcolor{deepcarrotorange}{orange}  in Figure \ref{fig:BBfusion}. This bounding box is fine-tuned by moving the boundary walls towards the centre by an offset value,  shown in \textbf{\textcolor{blue(ncs)}{blue}}  in Figure \ref{fig:BBfusion}. This offset factor is a percentage of the difference between the lowest and highest values of the lower bounds of individual bounding boxes. In Figure \ref{fig:iouplot},  a plot of the average IOU values against offset values is shown for each of the five fold cross validation results for localisation of the left and right kidneys. While the performance with zero offset is good with an IOU of 68\% (with values of IOU above 33\% (DC 50\%) considered acceptable), an optimal value is achieved for an offset value of 33\%. This value was found to hold good for lumbar spine and spleen as well and therefore was used as a criterion for fusing bounding boxes.
\subsubsection{Five fold cross validation results}
\hfill\\
In Tables \ref{table1} and \ref{table2}, the best Bounding box generated by all the models and the Fusion bounding box results for lumbar spine, spleen, left and right kidneys are shown, based on Accuracy, Jaccard Index (IOU), Dice Coefficient, and mean wall to wall and centroid to centroid distances.
\subsubsection{Three fold cross validation results}
\hfill\\
Three fold cross validation experiments were performed to establish a baseline for comparison with the semi supervised learning methods, for which three fold cross validation was performed.

\subsection{Semi Supervised Learning (SSL)}
Three different experiments were performed for localising the lumbar spine, spleen, left and right kidneys. 
\begin{enumerate}[label=(\alph*)]
    \item SSL 40:60 in which the entire data set was split randomly into 40\% labelled and the remaining unlabelled
    \item SSL 35:65 in which the entire data set was split randomly into 35\% labelled and the remaining unlabelled
    \item SSL 30:70 in which the entire data set was split randomly into 30\% labelled and the remaining unlabelled.
\end{enumerate}
The labelled and unlabelled sets were split into folds separately. During each fold of validation, data from the corresponding folds of labelled and unlabelled sets were combined to form a test set. Similarly the data from the remaining folds of labelled and unlabelled sets were combined to form the training set.

\begin{table}[!t]
\caption{\label{table1}Best bounding box organ fully supervised mode localisation results using five fold cross validation}
\centering
\begin{tabular}{@{}llllll}
\hline
Organ&Accuracy \%&Jaccard Index&Dice Coefficient&wl to  wl&ctd to ctd\\
&mn$\pm$SD&mn$\pm$SD&mn$\pm$SD&mn$\pm$SD&mn$\pm$SD\\
\hline
Lumbar Spine &97.27$\pm0.02$&71.19$\pm3.03$&82.55$\pm2.14$&7.83$\pm5$& 16.21$\pm13.41$\\
Spleen&89.48$\pm0.08$&62.25$\pm1.88$&76.05$\pm1.42$&7.33$\pm3.29$&12.08$\pm6.13$\\
Left Kidney&92.73$\pm0.05$&72.71$\pm3.29$&83.79$\pm2.18$&4.68$\pm2.122$&7.7$\pm4.36$\\
Right Kidney&99.05$\pm0.02$&74.51$\pm1$&85.06$\pm0.71$&4.38$\pm1.94$&6.84$\pm3.6$\\
\hline
\end{tabular}
\end{table}

\begin{table}[!t]
\caption{\label{table2}Fused bounding box organ fully supervised mode localisation results using five fold cross validation}
\centering
\begin{tabular}{@{}llllll}
\hline
Organ&Accuracy \%&Jaccard Index&Dice Coefficient&wl to  wl&ctd to ctd\\
&mn$\pm$SD&mn$\pm$SD&mn$\pm$SD&mn$\pm$SD&mn$\pm$SD\\
\hline
Lumbar Spine&97.27$\pm0.02$&66.86$\pm2.54$&79.52$\pm2.01$&9.46$\pm5.35$&20.14$\pm14.85$\\
Spleen&89.43$\pm0.05$&58.3$\pm1,56$&72.98$\pm1.18$&8.35$\pm3.39$&13.32$\pm6.74$\\
Left Kidney&90.91$\pm0.07$&69.54$\pm3$&81.57$\pm2.08$&5.23$\pm2.3$&8.4$\pm4.61$\\
Right Kidney&98.1$\pm0.03$&68.12$\pm1.38$&80.6$\pm0.95$&5.56$\pm2.34$&8.7$\pm5.12$\\
\hline
\end{tabular}
\end{table}

\begin{table}[!t]
\caption{\label{table3}Best bounding box Lumbar Spine semi supervised mode localisation results using three fold cross validation}
\centering
\begin{tabular}{@{}llllll}
\hline
Mode&Accuracy \%&Jaccard Index \%&Dice Coefficient \%&wl to wl&ctd to ctd\\ 
&mn$\pm$SD&mn$\pm$SD&mn$\pm$SD&mn$\pm$SD&mn$\pm$SD\\
\hline
Supervised&97.25$\pm0.03$&71.89$\pm1.81$&83.17$\pm1.15$&7.35$\pm4.75$&14.62$\pm12.9$\\
SSL 40:60&98.2$\pm0.02$&65.28$\pm3.6$&78.37$\pm2.75$&10.1$\pm5.79$&21.22$\pm15.9$\\
SSL 35:65&98.17$\pm0.02$&68.16$\pm0.11$&80.46$\pm0.13$&8.53$\pm4.54$&16.48$\pm12.01$\\
SSL 30:70&97.27$\pm0$&65.33$\pm0.88$&78.47$\pm0.8$&10.01$\pm5.04$&20.38$\pm14.25$\\
\hline
\end{tabular}
\end{table}
\begin{table}[!t]
\caption{\label{table4}Fused bounding box Lumbar spine semi supervised mode localisation results using three fold cross validation}
\centering
\begin{tabular}{@{}llllll}
\hline
Mode&Accuracy \%&Jaccard Index \%&Dice Coefficient \%&wl to wl&ctd to ctd\\ 
&mn$\pm$SD&mn$\pm$SD&mn$\pm$SD&mn$\pm$SD&mn$\pm$SD\\
\hline
Supervised&97.27$\pm0.02$&66.86$\pm2.54$&79.2$\pm2.01$&9.46$\pm5.35$&20.14$\pm14.85$\\
SSL 40:60&94.54$\pm0$&65.95$\pm2.65$&78.8$\pm0.03$&10.01$\pm$5.83&21.87$\pm15.75$\\
SSL 35:65&94.54$\pm0$&	64.1$\pm1.5$&77.44$\pm1.19$&11.47$\pm5.23$&24.05$\pm14.11$\\
SSL 30:70&96.35$\pm0.02$&61.26$\pm0.62$&75.2$\pm0.65$&11.47$\pm5.64$&24.05$\pm15.82$\\
\hline
\end{tabular}
\end{table}
\begin{table}[!t]
\caption{\label{table5}Best bounding box spleen semi supervised mode localisation results using three fold cross validation}
\centering
\begin{tabular}{@{}llllll}
\hline
Mode&Accuracy \%&Jaccard Index \%&Dice Coefficient \%&wl to wl&ctd to ctd\\ 
&mn$\pm$SD&mn$\pm$SD&mn$\pm$SD&mn$\pm$SD&mn$\pm$SD\\
\hline
Supervised&96.19$\pm0.04$&62.67$\pm1.25$&76.48$\pm0.96$&7.13$\pm3.02$&11.67$\pm5.86$\\
SSL 40:60&87.48$\pm0.06$&59.47$\pm0.15$&78.05$\pm0.16$&7.99$\pm2.79$&13.96$\pm5.83$ \\
SSL 35:65&87.56$\pm0.11$&58.2$\pm2.73$&73$\pm2.3$&8.28$\pm3.09$&13.45$\pm6.38$\\
SSL 30:70&85.61$\pm0.04$&58.23$\pm3.02$&72.87$\pm2.48$&8.38$\pm3.19$&13.62$\pm6.35$\\
\hline
\end{tabular}
\end{table}
\begin{table}[!t]
\caption{\label{table6}Fused bounding box spleen semi supervised mode localisation results using three fold cross validation}
\centering
\begin{tabular}{@{}llllll}
\hline
Mode&Accuracy \%&Jaccard Index \%&Dice Coefficient \%&wl to wl&ctd to ctd\\ 
&mn$\pm$SD&mn$\pm$SD&mn$\pm$SD&mn$\pm$SD&mn$\pm$SD\\
\hline
Supervised&92.35$\pm0.04$&57.79$\pm2.35$&72.63$\pm1.99$&8.39$\pm3.34$& 13.65$\pm6.49$\\
SSL 40:60&82.61$\pm0.09$&55.26$\pm0.96$&70.58$\pm0.91$&9.21$\pm3.2$&15.55$\pm7.22$\\
SSL 35:65&78.74$\pm0.14$&54.7$\pm2.28$&70.13$\pm1.96$&9.19$\pm3.27$&14.98$\pm6.8$\\
SSL 30:70&74.95$\pm0.12$&55.13$\pm2.06$&70.39$\pm1.69$&9.11$\pm3.42$&14.93$\pm6.39$\\
\hline
\end{tabular}
\end{table}
\begin{table}[!t]
\caption{\label{table7}Best bounding box left kidney semi supervised mode localisation results using three fold cross validation}
\centering
\begin{tabular}{@{}llllll}
\hline
Mode&Accuracy \%&Jaccard Index \%&Dice Coefficient \%&wl to wl&ctd to ctd\\ 
&mn$\pm$SD&mn$\pm$SD&mn$\pm$SD&mn$\pm$SD&mn$\pm$SD\\
\hline
Supervised&93.59$\pm0.06$&72.75$\pm1.05$&83.75$\pm0.85$&4.69$\pm2.35$&7.67$\pm4.75$\\
SSL 40:60&93.59$\pm0.04$&67.13$\pm1.64$&79.7$\pm1.33$&5.94$\pm2.96$&9.69$\pm6.57$\\
SSL 35:65&93.62$\pm0.06$&66.37$\pm0.97$&79.19$\pm0.82$&6.22$\pm2.93$&10.45$\pm6.57$\\
SSL 30:70&82.76$\pm0.06$&65.36$\pm3.63$&76.88$\pm2.81$&6.69$\pm3.02$&10.93$\pm6.71$\\
\hline
\end{tabular}
\end{table}
\begin{table}[!t]
\caption{\label{table8}Fused bounding box left kidney semi supervised mode localisation results using three fold cross validation}
\centering
\begin{tabular}{@{}llllll}
\hline
Mode&Accuracy \%&Jaccard Index \%&Dice Coefficient \%&wl to wl&ctd to ctd\\ 
&mn$\pm$SD&mn$\pm$SD&mn$\pm$SD&mn$\pm$SD&mn$\pm$SD\\
\hline
Supevised&89.99$\pm0.1$&68.15$\pm1.83$&80.52$\pm1.42$&5.55$\pm2.58$&8.86$\pm5.18$\\
SSL 40:60&85.39$\pm0.07$&63.71$\pm2.48$&77.13$\pm1.93$&6.55$\pm3.1$&10.69$\pm7$\\
SSL 35:65&89.04$\pm0.07$&63.66$\pm0.06$&77.18$\pm0.14$&6.69$\pm3.02$&11.57$\pm6.83$\\
SSL 30:70&74.62$\pm0.11$&61.76$\pm2.45$&75.69$\pm2.07$&6.96$\pm3.01$&11.18$\pm.65$\\
\hline
\end{tabular}
\end{table}
\begin{table}[!t]
\caption{\label{table9}Best bounding box right kidney semi supervised mode localisation results using three fold cross validation}
\centering
\begin{tabular}{@{}llllll}
\hline
Mode&Accuracy \%&Jaccard Index \%&Dice Coefficient \%&wl to wl&ctd to ctd\\ 
&mn$\pm$SD&mn$\pm$SD&mn$\pm$SD&mn$\pm$SD&mn$\pm$SD\\
\hline
Supervised&98.07$\pm0.02$&69.9$\pm2.49$&81.68$\pm1.91$&5.36$\pm2.81$&9.01$\pm5.48$\\
SSL 40:60&95.18$\pm0.02$&66.97$\pm1.05$&79.58$\pm0.77$&6$\pm2.83$&9.76$\pm55.76$\\
SSL 35:65&93.33$\pm0.07$&66.03$\pm0.98$&78.81$\pm1.03$&6.23$\pm3.02$&9.87$\pm6.04$\\
SSL 30:70&96.19$\pm0.04$&66.3$\pm3.19$&79.04$\pm2.37$&6.28$\pm3.07$&10.24$\pm5.98$\\
\hline
\end{tabular}
\end{table}
\begin{table}[!t]
\caption{\label{table10}Fused bounding box right kidney semi supervised mode localisation results using three fold cross validation}
\centering
\begin{tabular}{@{}llllll}
\hline
Mode&Accuracy \%&Jaccard Index \%&Dice Coefficient \%&wl to wl&ctd to ctd\\ 
&mn$\pm$SD&mn$\pm$SD&mn$\pm$SD&mn$\pm$SD&mn$\pm$SD\\
\hline
Supervised&98.07$\pm0.02$&64.65$\pm2.16$&77.86$\pm1.71$&6.41$\pm2.97$&10.57$\pm5.82$\\
SSL 40:60&92.27$\pm0.03$&61.97$\pm1.53$&75.9$\pm1.14$&7$\pm3$&11.38$\pm6.05$\\
SSL 35:65&91.4$\pm0.08$&60.74$\pm1.37$&74.71$\pm1.24$&7.19$\pm3.42$&11.79$\pm6.94$\\
SSL 30:70&94.26$\pm0.03$&61.09$\pm2.98$&74.98$\pm2.37$&7.23$\pm3.43$&11.98$\pm6.88$\\
\hline
\end{tabular}
\end{table}
In Tables \ref{table3}, \ref{table5}, \ref{table7} and \ref{table9} the performances of the best bounding boxes generated by the three models for lumbar spine, spleen, left and right kidneys respectively are shown. In Tables \ref{table4}, \ref{table6}, \ref{table8} and \ref{table10}  the performances of the fused bounding boxes are shown. From the results, the following may be summarised:
\begin{enumerate}[label = (\alph*)]
    \item The lumbar spine occupies a fairly large volume in abdominal scans, and Figure \ref{fig:Annotation} provides an example. Refer Tables \ref{table3} and \ref{table4}. As can be seen the best performance was achieved when the fully supervised mode is used, however nearly comparable performance is achieved using SSL methods. Better performance was achieved by the best bounding boxes compared to the fused bounding boxes, based on  the mean centroid to centroid distance. There are individual differences between SSL methods attributable to the variations in data in the annotated scans, and the performance depends on the ability of the annotated data to generalise the model performance.
    \item Tables \ref{table5} and \ref{table6} summarises the localisation results for spleen.  There is large variation in spleen sizes, which contributes to lower IOU and Dice coefficient values compared to other organs (ref Tables \ref{table3}, \ref{table4}, \ref{table7}, \ref{table8}, \ref{table9} and \ref{table10}). Fully supervised methods performed the best. The relatively better performance in SSL 30:70 may be attributed to the ability of the annotated data to generalise the model performance.
    \item Left Kidney localisation results are summarised in Tables \ref{table7} and \ref{table8}. Fully supervised methods  performed best for the Left Kidney, and   SSL 40:60 and SSL 35:65  performed comparably. SSL 30:70 did not perform well in comparison, which suggests that the chosen annotated data was insufficient to generalise the model.
    \item Fully supervised methods performed the best for the right kidney as can be seen in Tables \ref{table9} and \ref{table10}, with nearly comparable performance by the SSL methods.
\end{enumerate}
\subsection{Comparison with other methods}
The performance of some reported methods on localisation of spleen and kidneys is shown in Table \ref{table11}. The first three rows in the table show the performance using Regression Forests by \citep{Criminisi2013}. \citep{Ghesu2017} used reinforcement learning but required a large data set of 1438 scans for left kidney and 1432 for right kidney.  (\citep{Humpire2018}), (\citep{Hussain2017})  and (\citep{Lu2016}) used CNN based methods combining detection in 2D planes to detect 3D bounding boxes. Good performance was achieved by  Humpire - Mamani \textit{et al},  however they used a large dataset consisting of 1884 scans, 60\% of which was used for training, 20\% for validation and 20\% for testing. By far the best performance was achieved by (\citep{Xu2019}) using a modified Faster RCNN method, while (\citep{Navarro2020}) used deep reinforcement learning. Most of the reported results on lumbar spine localisation (e.g. \citep{Suzani2015}, \citep{Chen2015}) focused on the centroid of the individual vertebrae rather than the centroid of the entire lumbar spine as  in this work. Also, many works did not use consistent metrics or combined their  results with segmentation (\citep{Sekuboyina2017}, \citep{Janssens2018}). In the last two rows of Table \ref{table11}, the mean centroid to centroid distances of individual vertebrae of lumbar spine is presented. 
While it is not possible to directly compare the results, as they were all trained and tested on different data sets, they may be used as indicators of performance of the methods. The  experiments in this work report five fold and three cross validation results for supervised learning approaches and three fold cross validation for semi supervised learning approaches and therefore are  more reliable. The data set  used in this work is  smaller compared to the other CNN based methods, however the IOU achieved in this work is better compared to the pure deep reinforcement learning method of (Navarro  \textit{et al} 2020) using a data set of similar size, and  the distance measures are comparable.

Even though SSL has been utilised in areas such as thoracic disease detection \citep{Wang2020} and lung nodule detection \citep{LiZ2019}, there is no published literature on organ localisation. Therefore there is no benchmark to compare the results of the proposed method with.  The fact that nearly comparable results can be achieved with fewer annotations is very encouraging. 
\begin{table}[!t]
\centering
\caption{\label{table11} Comparison of other methods}
\begin{tabular}{@{}llllll}
\hline
Method&Organ&Data&IOU&wl to wl&ctd to ctd\\
\hline
\citep{Criminisi2013}&Left Kidney&Train-55,Test-45& - &13.6$\pm12.5$&\\
\citep{Criminisi2013}&Right Kidney&Train-55,Test-45& - &16.1$\pm15.5$&\\
\citep{Criminisi2013}&Spleen&Train-55,Test-45& - &20.7$\pm22.8$&\\
\citep{Humpire2018}&Left Kidney&1884&84&2.67$\pm7.18$&\\
\citep{Humpire2018}&Right Kidney&1884&85&3.03$\pm9.30$&\\
\citep{Humpire2018}&Spleen&1884&84&3.37$\pm8.46$&\\
\citep{Ghesu2017}&Left Kidney&1438& - & - &6.72$\pm 3.62$\\
\citep{Ghesu2017}&Right Kidney&1432& - & - &6.89$\pm3.65$\\
\citep{Hussain2017}&Left Kidney&Train-140,Test-60& - &6.19$\pm6.02$&7.71$\pm4.91$\\
\citep{Hussain2017}&Right Kidney&Train-140,Test-60& - &5.86$\pm6.40$&7.56$\pm4.1$\\
\citep{Lu2016}&Right Kidney&Train-450,est-49& - & - &7.8$\pm9.4$\\
\citep{Xu2019}&Left Kidney&Train-130, Test-70&75.29&4.31$\pm4.18$&5.59$\pm2.86$\\
\citep{Xu2019}&Right Kidney&Train-130,Test-70&76.46&3.89$\pm3.47$&6.19$\pm3.76$\\
\citep{Xu2019}&Spleen&Train-130,Test-70&70.01&6.28$\pm6.65$&9.95$\pm5.84$\\
\citep{Navarro2020}&Left Kidneyt&Train-70,Test-20&57&4.06$\pm4.98$&7.52$\pm9.02$\\
\citep{Navarro2020}&Right Kidney&Train-70,Test-20&60&2.96$\pm2.91$&5.69$\pm5.6$\\
\citep{Navarro2020}&Spleen&Train-70,Test-20&60&5.25$\pm7.23$&9.20$\pm12.03$\\
\citep{Suzani2015}&Lumbar Spine&224& - & - &20.3$\pm12.1$\\
\citep{Chen2015}&Lumbar Spine&302& - & - &8.42$\pm8.62$\\
\hline\\
\end{tabular}
\end{table}
\newblock
\newpage
\section{Discussion and Conclusion}

 The proposed method for organ and ROI localisation is based on a unique approach combining reinforcement learning with imitation learning. Six bounding boxes are generated using 3 CNN architectures. The network is also trained to predict the confidence levels associated with the generated 6 bounding boxes, which can be used to select candidate bounding boxes. These candidates are fused together using a novel method and performance is reported using average wall and centroid distances in addition to Jaccard Index and Dice Coefficient. Compared to pure deep reinforcement learning, the proposed approach leads an agent to the target centroid itself rather than stopping after crossing a threshold  level, and is therefore likely to result in better training.  While there may be variations in data sets, the method performs well with a smaller data set. 
 The method was further adapted for use in an SSL setting and three fold cross validation was performed for 4 organs with nearly comparable results to the fully supervised model. Considering the difficulties associated with obtaining annotated data, SSL proves to be a workable solution especially considering the importance  of organ/region of interest localisation in medical image analysis. Localisation need not be restricted to organs, and it can be applied to sub organs or an other regions of interest sought by physicians and surgeons. The method was applied on CT images in this work, however it is applicable to other imaging modalities as well. 

\bibliographystyle{unsrtnat}
\bibliography{references}  

\end{document}